\documentclass[12pt]{article}
\usepackage{epsfig}

\def\beq{\begin{equation}}
\def\eeq#1{\label{#1}\end{equation}}
\def\eeqn{\end{equation}}

\def\beqa{\begin{eqnarray}}
\def\eeqa#1{\label{#1}\end{eqnarray}}
\def\eeqan{\end{eqnarray}}

\let\bar=\overbar

\def\L{{\cal L}}

\def\Dslash{\not{\hbox{\kern-4pt $D$}}}
\def\dslash{\not{\hbox{\kern-2pt $\del$}}}

\def\msb{{\bar{\ssstyle M \kern -1pt S}}}

\def\Title#1{\begin{center} {\Large {\bf #1} } \end{center}}

\begin{document}

\Title{$\sin2\phi_2(=\alpha)$ from Belle}

\bigskip\bigskip


\begin{raggedright}  

{\it Won, Eunil (representing the Belle Collaboration)\index{Won, Eunil}\\
Department of Physics \\
Korea University\\
Seoul 136-701 Korea}
\bigskip\bigskip
\end{raggedright}

\begin{center} 
Presented at ``Flavor Physics and $CP$ Violation'' (FPCP),\\
16-18 May 2002, University of Pennsylvania, Philadelphia, USA
\end{center}

\section{Introduction}

 Recent measurements of the $CP$-violating parameter $\sin2\phi_1$
by Belle~\cite{sin2phi1} and BaBar~\cite{sin2beta} collaborations established
$CP$ violation in the neutral $B$ meson system that is consistent with
Kobayashi and Maskawa (KM) expectations~\cite{km}. 
Therefore, measurements of other
$CP$-violating parameters such as $\phi_2$ and $\phi_3$ 
provide important tests of the KM model. In this
note we describe a measurement of $CP$-violating asymmetries in the mode
$B^0\rightarrow \pi^+\pi^-$ which are sensitive to the parameter 
$\sin2\phi_2$.

The KM model predicts $CP$-violating asymmetries in the time-dependent rates
for $B^0$ and $\bar{B}^0$ decays to a common $CP$ eigenstates, 
$f_{CP}$~\cite{sanda}. When the $\Upsilon(4S)$ decays into a 
$B^0\bar{B}^0$ meson pair, the two mesons remain in a coherent $p$-wave state
until one of them decays. The decay of one of the $B$ mesons at time
$t_{tag}$ to a final state, $f_{tag}$, which distinguishes
between $B^0$ and $\bar{B}^0$, projects the accompanying $B$ meson
onto the opposite $b$-flavor at $\tau_{tag}$; this meson decays to 
$\pi^+\pi^-$ at time $t_{CP}$. The decay rate has a time dependence given 
by~\cite{rate}
\begin{eqnarray}
\label{eq:R_q}
P_{\pi\pi}^q(\Delta{t}) = 
\frac{e^{-|\Delta{t}|/{\tau_{B^0}}}}{4{\tau_{B^0}}}
\left[1 + q\cdot 
\left\{ S_{\pi\pi}\sin(\Delta m_d\Delta{t})   \right. \right. \nonumber \\
\left. \left.
   + A_{\pi\pi}\cos(\Delta m_d\Delta{t})
\right\}
\right],
\end{eqnarray}
where $\tau_{B^0}$ is the $B^0$ lifetime, $\Delta m_d$ is the mass 
difference between the two $B^0$ mass eigenstates, $\Delta t$ = $t_{CP} -
t_{tag}$, and the $b$-flavor charge $q$ = +1 (-1) when the tagging $B$
meson is a $B^0$ ($\bar{B}^0$). The $CP$-violating parameters $S_{\pi\pi}$
and $A_{\pi\pi}$ defined in Eq.~(\ref{eq:R_q}) are expressed by
\begin{equation}
S_{\pi\pi} = \frac{2Im\lambda}{|\lambda|^2+1}
~{\rm and}~A_{\pi\pi} = \frac{|\lambda|^2-1}{|\lambda|^2+1}
\end{equation}
where $\lambda$ is a complex parameter that depends on both $B^0\bar{B}^0$
mixing and on the amplitudes for $B^0$ and $\bar{B}^0$ decays to 
$\pi^+\pi^-$. In the Standard Model, to a good approximation, 
$|\lambda|$ is equal to the absolute value of the ratio of the $\bar{B}^0$
to $B^0$ decay amplitudes. Therefore $|\lambda| \neq 1$, or equivalently 
$A_{\pi\pi} \neq 0$, indicates direct $CP$ violation. 

 In the case of $B^0 \rightarrow (c\bar{c})K^0_{s}$ $CP$ eigenstate decays,
the $CP$-violating parameters are rather precisely expressed as 
$S_{(c\bar{c})K^0_s}$ = $\sin{2\phi_1}$ and 
$A_{(c\bar{c})K^0_s}$ = 0. This is due to the fact that the tree 
amplitude with $W$ emission dominates the $b \rightarrow s$ penguin amplitude
with associated $c\bar{c}$ production, which is small and has the same 
weak phase. For the $B^0 \rightarrow \pi^+\pi^-$ decay, we would have
$S_{\pi\pi}$ = $\sin{2\phi_2}$ and 
$A_{\pi\pi}$ = 0 if the $b \rightarrow u$ tree amplitude were dominant. 
The situation is complicated by the possibility of significant contributions 
from gluonic $b \rightarrow d$ penguin amplitudes that have a different weak
phase and additional strong phases~\cite{penguin}.
As a result, $S_{\pi\pi}$ may not be equal to $\sin{2\phi_2}$ and direct
$CP$ violation, $A_{\pi\pi}\neq$ 0, may occur.
One can quantify the contributions from penguin amplitude as
$\sin{2(\phi_2+\theta)}$ and determine $\theta$ using other decays.
However, it requires measurements of branching fractions
of $B^0$ $\rightarrow$ $\pi^0\pi^0$ and $\bar{B}^0$ $\rightarrow$ $\pi^0\pi^0$
separately, which is experimentally not within reach at 
this moment~\cite{pi0pi0}.

 This measurement of $CP$-violating parameters in $B^0 \rightarrow \pi^+\pi^-$
is based on a 41.8 fb$^{-1}$ data sample, 
which contains 44.1 million $B\bar{B}$
pairs, collected with the Belle detector at the KEKB asymmetric-energy
$e^+e^-$ (3.5 on 8 GeV) collider operating at the $\Upsilon(4S)$ resonance.

\section{Event Selection}

\begin{figure}[htb]
\begin{center}
\epsfig{file=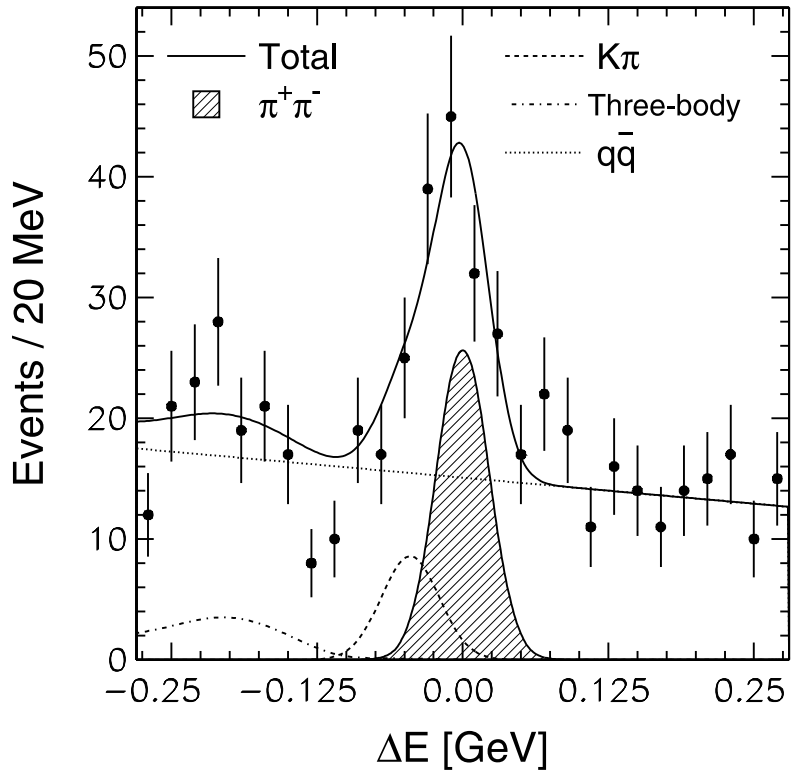,height=4in}
\caption{$\Delta E$ distribution for $\pi^+\pi^-$ event candidates that 
are in the $M_{bc}$ signal region.
Closed circles with error bars are the data, and individual curves are
results of fits using various components indicated in the figure.
}
\label{fig:de}
\end{center}
\end{figure}

 We use oppositely charged pairs of well measured tracks that
are positively 
identified as pions according to the combined information from the
ACC 
and the CDC $dE/dx$ measurement. Candidate $B$ mesons are reconstructed
using the energy difference $\Delta E$ $\equiv$ 
$E^{cms}_{B} - E^{cms}_{beam}$ and the beam energy constrained mass 
$M_{bc}$ $\equiv$ $\sqrt{(E^{cms}_{beam})^2 - (p^{cms}_{B})^2}$,
where $E^{cms}_{beam}$ is the cms beam energy, 
and $E^{cms}_{B}$ and $p^{cms}_{B}$ are the cms energy and momentum of
the $B$ candidate. The signal region is defined as
5.271 GeV/$c^2$ $<$ $M_{bc}$ $<$ 5.287 GeV/$c^2$ and $|\Delta E|$ 
$<$ 0.067 GeV, corresponding to $\pm$ 3$\sigma$ from the central
values. In order to suppress background from the $e^+e^-$ $\rightarrow$
$q\bar{q}$ continuum ($q$ = $u,d,s,c$), we form signal and background
likelihood functions, $\L_{S}$ and $\L_{BG}$, from two variables.
One is a Fisher discriminant determined from six modified Fox-Wolfram 
moments~\cite{fox} and the other is the $B$ flight direction in the
cms, with respect to the $z$ axis ($\cos{\theta_B}$). 
We determine
$\L_S$ from Monte Carlo (MC) and $\L_{BG}$ from data. The likelihood
ratio $\L_S/$ $(\L_S+\L_{BG})$ for candidate events is required to be 
greater than 0.825. Figure~\ref{fig:de} shows the $\Delta E$ distribution
for $\pi^+\pi^-$ candidates. The signal yield is extracted by fitting
the $\Delta E$ distribution with a Gaussian $\pi^+\pi^-$ signal function,
plus contributions from misidentified $B^0 \rightarrow K^+\pi^-$
events, three-body $B$-decays, and continuum background. From the fit,
we obtain 73.5 $\pm$ 13.8 events, 28.4 $\pm$ 12.5 
$K^+\pi^-$ events, and 98.7 $\pm$ 7.0 continuum background events in 
the signal region, where errors are statistical only. The $K^+\pi^-$
contamination level is consistent with the $K \rightarrow \pi$ 
misidentification probability measured independently, and the contribution
from three-body $B$-decays is little in the signal region.

 Leptons, charged pions, and kaons that are not associated with the 
reconstructed $B^0 \rightarrow \pi^+\pi^-$ decay are used to identify
the flavor of the accompanying $B$ meson. We apply the same method used for 
the $\sin{2\phi_1}$ measurement~\cite{sin2phi1}. We use two parameters,
$q$ and $r$, to represent the tagging information. The first, $q$,
corresponds to the sign of the $b$ quark charge where $q$ = +1 for $\bar{b}$
and hence $B^0$, and $q$ = $-$1 for $b$ and $\bar{B}^0$. The parameter
$r$ is an event-by-event, MC-determined flavor-tagging dilution factor that 
ranges from $r$ = 0 for no flavor discrimination to $r$ = 1 for unambiguous
flavor assignment. It is used only to sort data into six intervals of $r$,
according to flavor purity.

 The wrong tag fractions, $w_l$ ($l$=1,6), are determined directly from the 
data for the six $r$ intervals using exclusively reconstructed, self-tagged
$B^0 \rightarrow D^{*-}l^+\nu$,$D^{(*)}\pi^+$, and $D^{*-}\rho^+$ decays.
The $b$-flavor of the accompanying $B$ meson is assigned according to
the flavor-tagging algorithm described above. The decay vertices are
reconstructed using the same algorithm used for the accompanying 
$B$ mesons of $B^0 \rightarrow \pi^+\pi^-$ candidates. The values of
$w_l$ are obtained from the time evolution of neutral $B$ meson pairs 
with opposite flavor (OF) or same flavor (SF), which is given by 
$P_{OF[SF]}(\Delta t)$ = $e^{-|\Delta t|/\tau_{B^0}}$/$4\tau_{B^0}$
\{1+[$-$](1$-$2$w_l$)$\cos{(\Delta m_d \Delta t)}$\}. For the fits, we fix
$\Delta m_d$ at the world average value~\cite{pdg}.

 The vertex positions for the $\pi^+\pi^-$ and $f_{tag}$ decays are
reconstructed using tracks that have at least one three-dimensional 
coordinate determined from associated $r-\phi$ and $z$ hits in the
same SVD layer and one or more additional $z$ hits in the other layers.
Each vertex position is required to be consistent with the interaction point
profile smeared in the $r-\phi$ plane by the average transverse 
$B$ meson decay length. The $f_{tag}$ vertex is determined from all well
reconstructed tracks, excluding the $B^0 \rightarrow \pi^+\pi^-$ 
candidate. Tracks that form a $K^0_s$ candidate are not used. The MC
simulation indicates that the typical vertex-finding efficiency is 92\%
and 91\% for $\pi^+\pi^-$ and $f_{tag}$, respectively, while the typical
vertex rms resolution in the $z$ coordinate is 75 $\mu$m for $B^0 \rightarrow$
$\pi^+\pi^-$ decays and 140 $\mu$m for $f_{tag}$ decays. 

 The proper-time interval resolution for the signal, $R_{sig}(\Delta t)$,
is obtained by convolving a sum of two Gaussians with a
function that takes into account the cms motion of the $B$ mesons. The
fraction in the main Gaussian is determined to be 0.97 $\pm$ 0.02 from a 
study of $B^0 \rightarrow $ $D^{*-}\pi^+$, $D^{*-}\rho^+$,
$D^-\pi^+$, $J/\psi K^{*0}$, $J/\psi K^0_S$ and $B^+ \rightarrow$
$\bar{D}^0\pi^+$, $J/\psi K^+$ events. The means 
($\mu_{\rm main}$, $\mu_{\rm tail}$)
and widths
($\sigma_{\rm main}$, $\sigma_{\rm tail}$)
of the Gaussians are
calculated event-by-event from the
$f_{CP}$ and $f_{\rm tag}$ vertex fit error matrices
and the $\chi^2$ values of the fit;
typical values are $\mu_{\rm main}=-0.24~{\rm ps}$,
$\mu_{\rm tail}=0.18~{\rm ps}$
and
$\sigma_{\rm main}=1.49~{\rm ps}$, $\sigma_{\rm tail}=3.85~{\rm ps}$.
The background resolution function $R_{q\overline{q}}(\Delta t)$,
which is dominated by continuum background,
has the same functional
form but the parameters are obtained from a sideband region
in $M_{bc}$ and $\Delta E$.

\section{CP Fit Results}

We determine $CP$ violation parameters
by performing an
unbinned maximum-likelihood fit of a $CP$-violating
probability density function (pdf) to the 
observed $\Delta t$ distributions.
We define the likelihood value for each event as a
function of $S_{\pi\pi}$ and $A_{\pi\pi}$:
\begin{eqnarray}
P_i =
\int 
[\{f_{\pi\pi}^l P_{\pi\pi}^q(\Delta t^\prime, w_l;S_{\pi\pi}, 
A_{\pi\pi}) +
f_{K\pi}^l P_{K\pi}^q(\Delta t^\prime, w_l)\}
\cdot R_{sig}(\Delta t_i-\Delta t^\prime) 
\nonumber \\
+ f_{q\overline{q}}^l P_{q\overline{q}}(\Delta t^\prime)
\cdot R_{q\overline{q}}(\Delta t_i-\Delta t^\prime)]d\Delta t^\prime.
\label{eq:likelihood}
\end{eqnarray}
Here $f_{\pi\pi}^l$, $f_{K\pi}^l$, and 
$f_{q\overline{q}}^l$ ($= 1 - f_{\pi\pi}^l - f_{K\pi}^l$) are the fractions 
of $\pi^+\pi^-$ signal,  
$K^+\pi^-$ background, and continuum background in flavor-tagging interval
$l$, respectively.  
These fractions are determined on an event-by-event basis
as a function of $\Delta E$ and $M_{\rm bc}$, properly normalized
by the average signal and background fractions in the
signal region.
For higher $r$ values where we are more sensitive to the
asymmetry, the fraction of continuum background decreases;
the ratio of $\pi^+\pi^-$ signal events to background
$K^+\pi^-$ events is the same for all $r$ bins.
The pdfs for $\pi^+\pi^-$ ($P_{\pi\pi}^q$), 
$K^+\pi^-$ ($P_{K\pi}^q$), and continuum background 
($P_{q\overline{q}}$),
are convolved with their respective resolution functions. 
We use the same vertex resolution 
function for $\pi^+\pi^-$ and $K^+\pi^-$ candidates.
For the $\pi^+\pi^-$ signal,
the pdf is given by Eq.~(\ref{eq:R_q}) with $q$ replaced by
$q(1-2w_l)$, to account for the dilution due to
wrong flavor tagging.
The pdf for the $K^+\pi^-$ background is
$P_{K\pi}^q(\Delta t,w_l)
={e^{-|\Delta t|/\tau_{B^0}}}/{4\tau_{B^0}}
\{ 1 + q\cdot(1-2w_l)
  A_{K\pi}\cos(\Delta m_d\Delta{t}) \}$,
where $A_{K\pi}$ is the $\bar{B}^0 \rightarrow K^-\pi^+$ and
$B^0 \to K^+\pi^-$ decay rate asymmetry.
We fix $A_{K\pi} = 0$ and $\tau_{B^0}$ and $\Delta m_d$ to their world average
values~\cite{pdg}.
Although the $K^+\pi^-$ background contamination is low, a possible
deviation of $A_{K\pi}$ from zero is included in the
systematic error. The pdf used for the $q\bar{q}$ background
distribution is $P_{q\bar{q}}(\Delta t)$ = \{$f_{\tau}$
$e^{-|\Delta t|/\tau_{bkg}}$/$2\tau_{bkg}$+(1-$f_{\tau}$)$\delta(\Delta t)$
\}/2,
where $f_{\tau}$ is the background fraction with an effective lifetime
$\tau_{bkg}$ and $\delta$ is the Dirac delta function. We determine
$f_{\tau}$ = 0.011 $\pm$ 0.004 and $\tau_{bkg}$ = 2.7$^{+1.0}_{-0.7}$ 
ps from the sideband data. In function $\L$ = $\Pi P_i$, where
the product is over all $B^0 \rightarrow$ $\pi^+\pi^-$
candidates.

\begin{figure}[htb]
\begin{center}
\epsfig{file=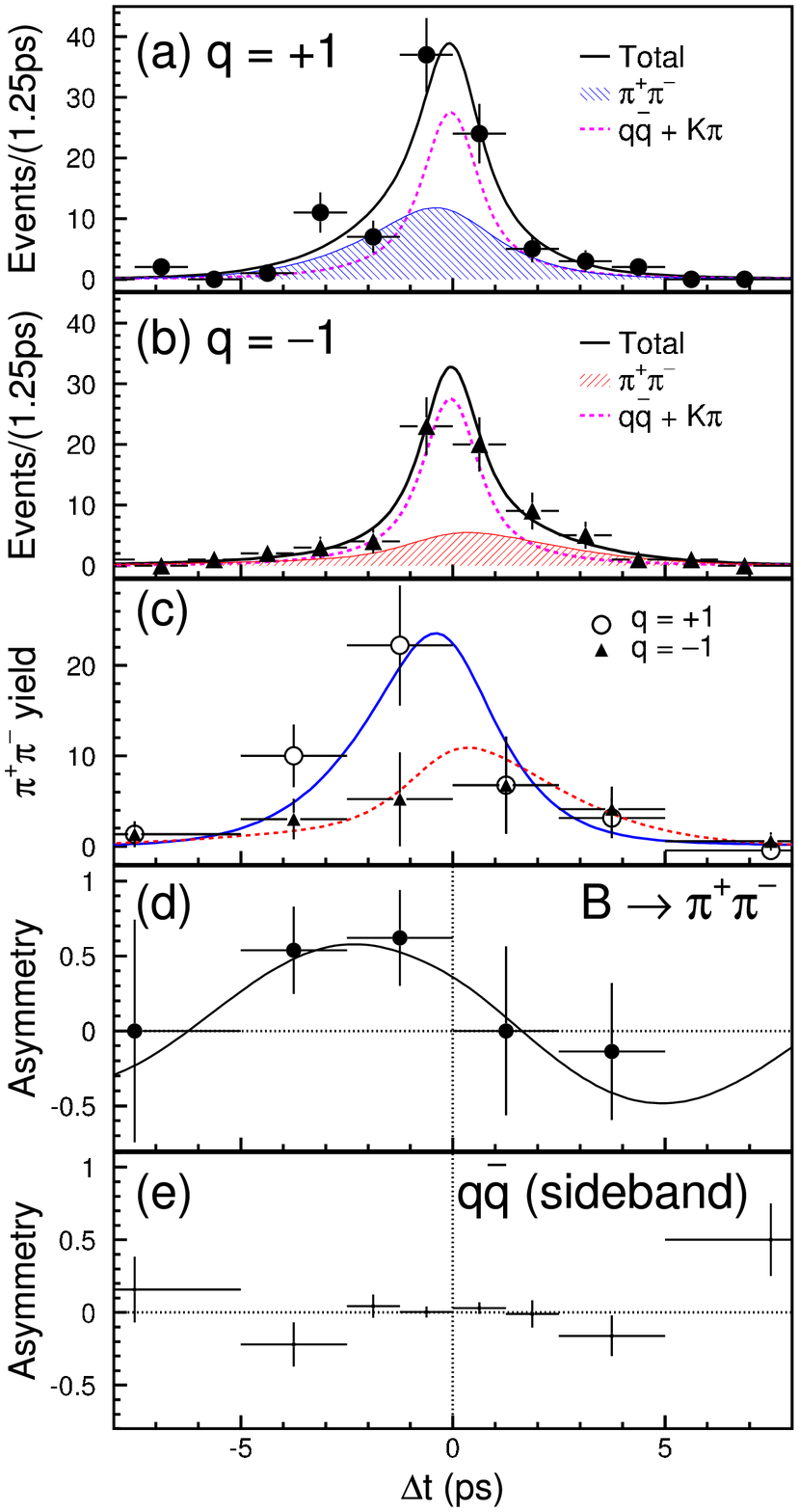,height=6in}
\caption{
The $\Delta t$ distributions 
 for the $B^0 \rightarrow \pi^+\pi^-$ candidates in the
signal region: 
(a) candidates with $q = +1$, i.e. the tag side is identified as $B^0$;
(b) candidates with $q = -1$; 
(c) $\pi^+ \pi^-$ yields after background subtraction. The rightmost
(leftmost) bin ranges from 5 to 10 ps ($-5$ to $-10$ ps);
(d) the $CP$ asymmetry for $B^0 \rightarrow \pi^+\pi^-$
after background subtraction. The
point in the rightmost bin has a large negative value
that is outside of the range of the histogram;
(e) the raw asymmetry for $B^0 \to \pi^+\pi^-$ sideband events.
In Figs. (a) through (d), the curves show the
results of the unbinned maximum likelihood fit.
}
\label{fig:fit}
\end{center}
\end{figure}

 The result of the fit to the 162 candidates (92 $B^0$- and 70 
$\bar{B}^0$-tags) that remain after flavor tagging and vertex reconstruction 
is:
\begin{eqnarray}
S_{\pi\pi} &=& -1.21~{^{+0.38}_{-0.27}}{\rm (stat)}~{^{+0.16}_{-0.13}}{\rm (syst)};
 \nonumber \\
A_{\pi\pi} &=& +0.94~{^{+0.25}_{-0.31}}{\rm (stat)}~{\pm 0.09}{\rm (syst)}.
 \nonumber
\end{eqnarray}
In Figs.~\ref{fig:fit}(a) and (b), we show the $\Delta t$ distributions
for $B^0$- and $\bar{B}^0$-tagged events together with the fit curves;
the background-subtracted $\Delta t$ distributions are shown in 
Fig.~\ref{fig:fit}(c). 
It appears that there are more $B^0$ tags than $\bar{B}^0$ tags.
Figure~\ref{fig:fit}(d) shows the background-subtracted
$CP$ asymmetry between the $B^0$- and $\bar{B}^0$-tagged events as a 
function of $\Delta t$, with the result of the fit superimposed.

\section{Systematic Uncertainties}

The systematic error on $S_{\pi\pi}$ is primarily due to uncertainties in 
the background fractions ($\pm$0.09) and from the fit bias near the
physical boundary ($^{+0.11}_{-0.02}$). For $A_{\pi\pi}$, the background
fractions ($\pm$0.06) and the wrong-tag fractions ($\pm$0.06)
are the two leading components. Other sources of systematic error are
uncertainties in the resolution function, physics parameters
($\Delta m_d$, $\tau_{B^0}$, and $A_{K\pi}$) and the background modeling.
A value of $A_{K\pi}$ = $-$0.06 $\pm$ 0.08 is obtained from the
self-tagged $B^0 \rightarrow K^+\pi^-$ sample. This introduces a systematic
error of $<$ 0.01 for $S_{\pi\pi}$ and $^{+0.02}_{-0.01}$ for $A_{\pi\pi}$.

We perform a number of cross checks. We examine the event yields and 
$\Delta t$ distributions for $B^0$- and $\bar{B}^0$-tagged events in 
the sideband region and find no significant difference between 
the two samples. Figure~\ref{fig:fit}(e) shows the observed raw asymmetry in 
the side band region. We select $B^0 \rightarrow K^+\pi^-$ candidates,
which have the same track topology as
$B^0 \rightarrow \pi^+\pi^-$, by positively identifying charged kaons. A 
fit to 309 events yields $A_{K\pi}$ = 0.07 $\pm$ 0.17, consistent with
the value mentioned above, and $S_{K\pi}$ 0.15 $\pm$ 0.24. We also select
$B^0 \rightarrow$ $D^-\pi^+$, $D^{*-}\pi^+$ and $D^-\rho^+$ candidates
using the same event shape criteria. These are non-$CP$ eigenstate
self-tagged modes, where neither mixing-induced nor direct $CP$-violating
asymmetry is expected. We find no asymmetry in these modes. As an 
additional test of the consistency of the background treatment, we
add events from the $B^0 \rightarrow \pi^+\pi^-$ sideband and adjust their
$\Delta E$ and $M_{bc}$ values. A fit to this background-enriched control 
sample, which has a similar background fraction as the 
$B^0 \rightarrow \pi^+\pi^-$ sample, yields $S$ = 0.08 $\pm$ 0.06 and
$A$ = 0.03 $\pm$ 0.04, both consistent with a null asymmetry.
We measured $\tau_{B^0}$ from $B^0$ $\rightarrow$ $\pi^+\pi-$ and
$K^+\pi^-$ decays and obtained 1.49 $\pm$ 0.21 ps and 1.73 $\pm$ 0.15 ps,
respectively (The world average value is 1.548 $\pm$ 0.032 ps~\cite{pdg}).
We measured $\Delta m_d$ from $B^0$ $\rightarrow$ $K^+\pi^-$ and 
found $\Delta m_d$ to be 0.57 $\pm$ 0.08 ps$^{-1}$, which is consistent
with the current world average value (0.472 $\pm$ 0.017 
ps$^{-1}$~\cite{pdg}).

 Our central values of $S_{\pi\pi}$ and $A_{\pi\pi}$ are 1.2$\sigma$ 
away from the physical boundary ($S_{\pi\pi}^2$ + $A_{\pi\pi}^2$ $<$ 1).
However, the components of the present analysis are the same as those used to
measure $\sin{2\phi_1}$, $\tau_{B}$, and $\Delta m_d$, which are all
in good agreement with the world average values and 
we performed ensemble tests and found no indication of bias in the
procedure and the likelihood fit errors are in reasonable agreement 
with expected from ensemble tests. Therefore, 
we attribute our central values of $S_{\pi\pi}$ and  $A_{\pi\pi}$
being outside of the physical boundary as a statistical fluctuation.

\section{Conclusions}

 In summary, we have measured the $CP$ violation parameters in $B^0$ 
$\rightarrow \pi^+\pi^-$ decay. Our results for $S_{\pi\pi}$ 
indicates that mixing-induced $CP$ violation is large. The large
$A_{\pi\pi}$ term is an indication of direct $CP$ violation in 
$B$ meson decay, and suggests that there is a large hardronic phase
and interference between the tree and penguin amplitudes. In this case,
the precise determination of $\sin{2\phi_2}$ from $S_{\pi\pi}$ 
requires additional measurements including the branching fractions for
the decays $B^0 \rightarrow$ $\pi^0\pi^0$ and $\bar{B}^0 \rightarrow$
$\pi^0\pi^0$~\cite{pi0pi0}, which will be performed in 
the near future at $B$ factories.

\end{document}